\documentclass[10pt,a4paper,onecolumn]{article}
\usepackage[T1]{fontenc}
\usepackage{lmodern}
\usepackage{marginnote}
\usepackage{graphicx}
\usepackage{xcolor}
\usepackage{authblk,etoolbox}
\usepackage{titlesec}
\usepackage{calc}
\usepackage{tikz}
\usepackage{hyperref}
\hypersetup{colorlinks,breaklinks,
            urlcolor=[rgb]{0.0, 0.5, 1.0},
            linkcolor=[rgb]{0.0, 0.5, 1.0}}
\usepackage{caption}
\usepackage{tcolorbox}
\usepackage{amssymb,amsmath}
\usepackage{ifxetex,ifluatex}
\usepackage{seqsplit}
\usepackage{xstring}

\usepackage{float}
\let\origfigure\figure
\let\endorigfigure\endfigure
\renewenvironment{figure}[1][2] {
    \expandafter\origfigure\expandafter[H]
} {
    \endorigfigure
}

\usepackage{fixltx2e} % provides \textsubscript
\usepackage[
  backend=biber,
]{biblatex}
\bibliography{paper.bib}

\let\textttOrig=\texttt
\def\texttt#1{\expandafter\textttOrig{\seqsplit{#1}}}
\renewcommand{\seqinsert}{\ifmmode
  \allowbreak
  \else\penalty6000\hspace{0pt plus 0.02em}\fi}

\makeatletter
\let\href@Orig=\href
\def\href@Urllike#1#2{\href@Orig{#1}{\begingroup
    \def\Url@String{#2}\Url@FormatString
    \endgroup}}
\def\href@Notdoi#1#2{\def\tempa{#1}\def\tempb{#2}%
  \ifx\tempa\tempb\relax\href@Urllike{#1}{#2}\else
  \href@Orig{#1}{#2}\fi}
\def\href#1#2{%
  \IfBeginWith{#1}{https://doi.org}%
  {\href@Urllike{#1}{#2}}{\href@Notdoi{#1}{#2}}}
\makeatother

\usepackage[top=3.5cm, bottom=3cm, right=1.5cm, left=1.0cm,
            headheight=2.2cm, reversemp, includemp, marginparwidth=4.5cm]{geometry}

\titleformat{\section}
  {\normalfont\sffamily\Large\bfseries}
  {}{0pt}{}
\titleformat{\subsection}
  {\normalfont\sffamily\large\bfseries}
  {}{0pt}{}
\titleformat{\subsubsection}
  {\normalfont\sffamily\bfseries}
  {}{0pt}{}
\titleformat*{\paragraph}
  {\sffamily\normalsize}

\usepackage{fancyhdr}
\makeatletter
\let\ps@plain\ps@fancy
\fancyheadoffset[L]{4.5cm}
\fancyfootoffset[L]{4.5cm}

\definecolor{linky}{rgb}{0.0, 0.5, 1.0}

\newtcolorbox{repobox}
   {colback=red, colframe=red!75!black,
     boxrule=0.5pt, arc=2pt, left=6pt, right=6pt, top=3pt, bottom=3pt}

\newcommand{\ExternalLink}{%
   \tikz[x=1.2ex, y=1.2ex, baseline=-0.05ex]{%
       \begin{scope}[x=1ex, y=1ex]
           \clip (-0.1,-0.1)
               --++ (-0, 1.2)
               --++ (0.6, 0)
               --++ (0, -0.6)
               --++ (0.6, 0)
               --++ (0, -1);
           \path[draw,
               line width = 0.5,
               rounded corners=0.5]
               (0,0) rectangle (1,1);
       \end{scope}
       \path[draw, line width = 0.5] (0.5, 0.5)
           -- (1, 1);
       \path[draw, line width = 0.5] (0.6, 1)
           -- (1, 1) -- (1, 0.6);
       }
   }

\patchcmd{\@maketitle}{center}{flushleft}{}{}
\patchcmd{\@maketitle}{center}{flushleft}{}{}
\patchcmd{\@maketitle}{\LARGE}{\LARGE\sffamily}{}{}
\def\maketitle{{%
  
  \AB@maketitle}}
\makeatletter
\renewcommand\AB@affilsepx{ \protect\Affilfont}
\renewcommand\AB@affilnote[1]{{\bfseries #1}\hspace{3pt}}
\renewcommand{\affil}[2][]%
   {\newaffiltrue\let\AB@blk@and\AB@pand
      \if\relax#1\relax\def\AB@note{\AB@thenote}\else\def\AB@note{#1}%
        \setcounter{Maxaffil}{0}\fi
        \begingroup
        \let\href=\href@Orig
        \let\texttt=\textttOrig
        \let\protect\@unexpandable@protect
        \def\thanks{\protect\thanks}\def\footnote{\protect\footnote}%
        \@temptokena=\expandafter{\AB@authors}%
        {\def\\{\protect\\\protect\Affilfont}\xdef\AB@temp{#2}}%
         \xdef\AB@authors{\the\@temptokena\AB@las\AB@au@str
         \protect\\[\affilsep]\protect\Affilfont\AB@temp}%
         \gdef\AB@las{}\gdef\AB@au@str{}%
        {\def\\{, \ignorespaces}\xdef\AB@temp{#2}}%
        \@temptokena=\expandafter{\AB@affillist}%
        \xdef\AB@affillist{\the\@temptokena \AB@affilsep
          \AB@affilnote{\AB@note}\protect\Affilfont\AB@temp}%
      \endgroup
       \let\AB@affilsep\AB@affilsepx
}
\makeatother

\renewcommand\Affilfont{\sffamily\small\mdseries}
\ifnum 0\ifxetex 1\fi\ifluatex 1\fi=0 % if pdftex
  \usepackage[T1]{fontenc}
  \usepackage[utf8]{inputenc}

\else % if luatex or xelatex
  \ifxetex
    \usepackage{mathspec}
  \else
    \usepackage{fontspec}
  \fi
  \defaultfontfeatures{Ligatures=TeX,Scale=MatchLowercase}

\fi
\IfFileExists{upquote.sty}{\usepackage{upquote}}{}
\IfFileExists{microtype.sty}{%
\usepackage{microtype}
\UseMicrotypeSet[protrusion]{basicmath} % disable protrusion for tt fonts
}{}

\usepackage{hyperref}
\hypersetup{unicode=true,
            pdftitle={UltraNest - a robust, general purpose Bayesian inference engine},
            pdfborder={0 0 0},
            breaklinks=true}
\let\addcontentslineOrig=\addcontentsline
\def\addcontentsline#1#2#3{\bgroup
  \let\texttt=\textttOrig\addcontentslineOrig{#1}{#2}{#3}\egroup}
\let\markbothOrig\markboth
\def\markboth#1#2{\bgroup
  \let\texttt=\textttOrig\markbothOrig{#1}{#2}\egroup}
\let\markrightOrig\markright
\def\markright#1{\bgroup
  \let\texttt=\textttOrig\markrightOrig{#1}\egroup}

\usepackage{graphicx,grffile}
\makeatletter
\def\maxwidth{\ifdim\Gin@nat@width>\linewidth\linewidth\else\Gin@nat@width\fi}
\def\maxheight{\ifdim\Gin@nat@height>\textheight\textheight\else\Gin@nat@height\fi}
\makeatother
\setkeys{Gin}{width=\maxwidth,height=\maxheight,keepaspectratio}
\IfFileExists{parskip.sty}{%
\usepackage{parskip}
}{% else
\setlength{\parindent}{0pt}
\setlength{\parskip}{6pt plus 2pt minus 1pt}
}
\providecommand{\tightlist}{%
  \setlength{\itemsep}{0pt}\setlength{\parskip}{0pt}}
\ifx\paragraph\undefined\else
\let\oldparagraph\paragraph
\renewcommand{\paragraph}[1]{\oldparagraph{#1}\mbox{}}
\fi
\ifx\subparagraph\undefined\else
\let\oldsubparagraph\subparagraph
\renewcommand{\subparagraph}[1]{\oldsubparagraph{#1}\mbox{}}
\fi

\title{UltraNest - a robust, general purpose Bayesian inference engine}

        \author[1, 2, 3, 4]{Johannes Buchner}
    
      \affil[1]{Max Planck Institute for Extraterrestrial Physics, Giessenbachstrasse,
85741 Garching, Germany.}
      \affil[2]{Millenium Institute of Astrophysics, Vicuña MacKenna 4860, 7820436
Macul, Santiago, Chile}
      \affil[3]{Pontificia Universidad Católica de Chile, Instituto de Astrofísica,
Casilla 306, Santiago 22, Chile.}
      \affil[4]{Excellence Cluster Universe, Boltzmannstr. 2, D-85748, Garching, Germany}
  \date{\vspace{-5ex}}

\begin{document}
\maketitle

\marginpar{
  \sffamily\small

  {\bfseries DOI:} \href{https://doi.org/}{\color{linky}{}}

  \vspace{2mm}

  {\bfseries Software}
  \begin{itemize}
    \setlength\itemsep{0em}
    \item \href{https://github.com/openjournals/joss-reviews/issues/}{\color{linky}{Review}} \ExternalLink
    \item \href{}{\color{linky}{Repository}} \ExternalLink
    \item \href{}{\color{linky}{Archive}} \ExternalLink
  \end{itemize}

  \vspace{2mm}

  {\bfseries Submitted:} \\
  {\bfseries Published:} 

  \vspace{2mm}
  {\bfseries License}\\
  Authors of papers retain copyright and release the work under a Creative Commons Attribution 4.0 International License (\href{https://creativecommons.org/licenses/by/4.0/}{\color{linky}{CC BY 4.0}}).
}

\hypertarget{summary}{%
\section{Summary}\label{summary}}

UltraNest is a general-purpose Bayesian inference package for parameter
estimation and model comparison. It allows fitting arbitrary models
specified as likelihood functions written in Python, C, C++, Fortran,
Julia or R. With a focus on correctness and speed (in that order),
UltraNest is especially useful for multi-modal or non-Gaussian parameter
spaces, computational expensive models, in robust pipelines.
Parallelisation to computing clusters and resuming incomplete runs is
available.

\hypertarget{statement-of-need}{%
\section{Statement of need}\label{statement-of-need}}

When scientific models are compared to data, two tasks are important: 1)
constraining the model parameters and 2) comparing the model to other
models. While several open source Bayesian model fitting packages are
available that can be easily tied to existing models, they are difficult
to run such that the result is reliable and user interaction is
minimized. A chicken-and-egg problem is that one does not know a priori
the posterior distribution of a given likelihood, prior and data set,
and cannot choose a sampler that performs well. For example, Markov
Chain Monte Carlo convergence checks may suggest good results, while in
fact another distant but important posterior peak has remained unseen.
Current and upcoming large astronomical surveys require characterising a
large number of highly diverse objects, which requires reliable analysis
pipelines. This is what UltraNest was developed for.

Nested sampling (NS, Skilling 2004) allows Bayesian inference on
arbitrary user-defined likelihoods. Additional to computing parameter
posterior samples, it also estimates the marginal likelihood
(``evidence'', \(Z\)). Bayes factors between two competing models
\(B=Z_1/Z_2\) are a measure of the relative prediction parsimony of the
models, and form the basis of Bayesian model comparison. By performing a
global scan of the parameter space from the worst to best fits to the
data, NS also performs well in multi-modal settings.

In the last decade, several variants of NS have been developed. The
variants relate to (1) how better and better fits are found while
respecting the priors, (2) whether it is allowed to go back to worse
fits and explore the parameter space more, and (3) diagnostics through
tests and visualisations. UltraNest develops novel, state-of-the-art
techniques for all of the above. They are especially remarkable for
being free of tuning parameters and theoretically justified.

Currently available efficient NS implementations such as
\texttt{MultiNest} (Feroz, Hobson, and Bridges 2009) and its open-source
implementations rely on a heuristic algorithm which has shown biases
when the likelihood contours are not ellipsoidal (Buchner 2014; Nelson
et al. 2020). UltraNest instead implements better motivated
self-diagnosing algorithms, and improved, conservative uncertainty
propagation. In other words, UltraNest prioritizes robustness and
correctness, and maximizes speed second. For potentially complex
posteriors where the user is willing to invest computation for obtaining
a gold-standard exploration of the entire posterior distribution in one
run, UltraNest was developed.

\begin{figure}
\centering
\includegraphics{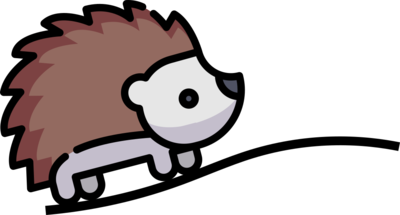}
\caption{Logo of UltraNest; made by
\url{https://www.flaticon.com/authors/freepik}}
\end{figure}

\hypertarget{method}{%
\section{Method}\label{method}}

NS methods are systematically reviewed in Buchner et al., submitted. The
approaches used in UltraNest are highlighted there as well.

The basic outline of vanilla NS (see Skilling 2004 for details) is as
follows:

A set of \(N\) points are randomly drawn from the unit hypercube
(u-space). A inverse cumulative prior transform converts these points to
physical parameter units (v-space). The likelihood \(L\) evaluated for
each point. In this population of live points, NS repeatedly replaces
the current worst likelihood point through likelihood-constrained prior
sampling (LRPS). At each iteration (represented by the removed, dead
point), the prior space investigated shrinks by approximately
\(V_{i+1}/V_i = (N - 1)/N\), starting from the entire prior volume
\(V_i=1\). The dead point becomes a posterior sample with weight
\(w_i=L_i\times V_i\), yielding the posterior distribution and the
evidence estimate \(Z_i=\sum_{j=1}^i w_i\). The iteration procedure can
terminate when the live points become unimportant, i.e.~when
\(w_{live}=V_{i+1}\max_{i=1}^N L_{live,i}\ll Z_i\).

\hypertarget{reactive-ns}{%
\subsection{Reactive NS}\label{reactive-ns}}

Instead of iterating with a fixed array of live points, UltraNest uses a
tree. The root of the tree represents the entire prior volume, and its
child nodes are samples from the entire prior. A breadth-first search is
run, which keeps a stack of the opened nodes sorted by likelihood. When
encountering a node, attaching a child to it is decided by several
criteria.

Reactive NS is a flexible generalisation of the Dynamic NS (Higson et
al. 2019), which used a simple heuristic for identifying where to add
more live points. The tree formulation of Reactive NS makes implementing
error propagation and variable number of live points straight-forward.

\hypertarget{integration-procedure}{%
\subsection{Integration procedure}\label{integration-procedure}}

UltraNest computes conservative uncertainties on \(Z\) and the posterior
weights. Several Reactive NS explorers are run which see only parts of
the tree, specifically a bootstrapped subsample of the root edges. For
each sample, each explorer estimates a weight (0 if it is blind to it),
and an estimate of the evidence. The ensemble gives an uncertainty
distribution.

The bootstrapped integrators are an evolution over single-bulk evidence
uncertainty measures and includes the scatter in likelihoods (by
bootstrapping) and volume estimates (by beta sampling; Skilling 2004).

\hypertarget{lrps-procedures-in-ultranest}{%
\subsection{LRPS procedures in
UltraNest}\label{lrps-procedures-in-ultranest}}

The live points all fulfill the current likelihood threshold, therefore
they can be used to trace out the neighbourhood where a new, independent
prior sample can be generated that also fulfills the threshold.
Region-based sampling uses rejection sampling using constructed
geometries.

UltraNest combines three region constructions, and uses their
intersection: 1) MLFriends (Buchner 2019), based on RadFriends (Buchner
2014), 2) a bootstrapped single ellipsoid in u-space and 3) another
bootstrapped single ellipsoid in v-space. The last one drastically helps
when one parameter constraint scales with another, (e.g., funnels).
UltraNest dynamically chooses whether to draw samples from the entire
prior, the single u-space ellipsoid or MLFriends ellipsoids (accounting
for overlaps), and filtered by the other constraints (including the
transformed v-space ellipsoid).

Useful for high dimensional problems (\(d>20\)), UltraNest supports
several types of Monte Carlo random walks, including:

\begin{itemize}
\tightlist
\item
  Slice sampling (as in \texttt{Polychord}, Handley, Hobson, and Lasenby
  2015)
\item
  Hit-and-run sampling
\item
  Constrained Hamiltonian Monte Carlo with No-U turn sampling (similar
  to \texttt{NoGUTS}, Griffiths and Wales 2019)
\end{itemize}

\hypertarget{features}{%
\section{Features}\label{features}}

\begin{itemize}
\tightlist
\item
  Run-time visualisation
\item
  Posterior visualisations
\item
  Diagnostic test of run quality
\item
  MPI parallelisation
\item
  Resuming
\item
  Models written in Python, C, C++, Fortran, Julia (Schulz and Buchner
  2020), R, and Javascript (Buchner 2018).
\end{itemize}

\href{https://johannesbuchner.github.io/UltraNest/}{Extensive
documentation} is available.

\hypertarget{acknowledgements}{%
\section{Acknowledgements}\label{acknowledgements}}

I am very thankful to Fred Beaujean, Josh Speagle and J. Michael Burgess
for insightful conversations.

\hypertarget{references}{%
\section*{References}\label{references}}
\addcontentsline{toc}{section}{References}

\hypertarget{refs}{}
\leavevmode\hypertarget{ref-Buchner2014stats}{}%
Buchner, Johannes. 2014. ``A statistical test for Nested Sampling
algorithms.'' \emph{Statistics and Computing}, July, 1--10.
\url{https://doi.org/10.1007/s11222-014-9512-y}.

\leavevmode\hypertarget{ref-ultranestjs}{}%
---------. 2018. ``Ultranest-Js.'' \emph{GitHub Repository}. GitHub.
\url{https://github.com/JohannesBuchner/ultranest-js}.

\leavevmode\hypertarget{ref-Buchner2019c}{}%
Buchner, Johannes. 2019. ``Collaborative Nested Sampling: Big Data
versus Complex Physical Models'' 131 (1004): 108005.
\url{https://doi.org/10.1088/1538-3873/aae7fc}.

\leavevmode\hypertarget{ref-Feroz2009}{}%
Feroz, F., M. P. Hobson, and M. Bridges. 2009. ``MULTINEST: an efficient
and robust Bayesian inference tool for cosmology and particle physics''
398 (October): 1601--14.
\url{https://doi.org/10.1111/j.1365-2966.2009.14548.x}.

\leavevmode\hypertarget{ref-Griffiths2019}{}%
Griffiths, Matthew, and David J Wales. 2019. ``Nested Basin-Sampling.''
\emph{Journal of Chemical Theory and Computation} 15 (12): 6865--81.
\url{https://doi.org/10.1021/acs.jctc.9b00567}.

\leavevmode\hypertarget{ref-Handley2015a}{}%
Handley, W. J., M. P. Hobson, and A. N. Lasenby. 2015. ``POLYCHORD:
next-generation nested sampling'' 453 (November): 4384--98.
\url{https://doi.org/10.1093/mnras/stv1911}.

\leavevmode\hypertarget{ref-Higson2017}{}%
Higson, Edward, Will Handley, Mike Hobson, and Anthony Lasenby. 2019.
``Dynamic nested sampling: an improved algorithm for parameter
estimation and evidence calculation.'' \emph{Statistics and Computing}
29 (5): 891--913. \url{https://doi.org/10.1007/s11222-018-9844-0}.

\leavevmode\hypertarget{ref-Nelson2020}{}%
Nelson, Benjamin E., Eric B. Ford, Johannes Buchner, Ryan Cloutier,
Rodrigo F. D\'iaz, João P. Faria, Nathan C. Hara, Vinesh M. Rajpaul, and
Surangkhana Rukdee. 2020. ``Quantifying the Bayesian Evidence for a
Planet in Radial Velocity Data'' 159 (2): 73.
\url{https://doi.org/10.3847/1538-3881/ab5190}.

\leavevmode\hypertarget{ref-ultranestjl}{}%
Schulz, Oliver, and Johannes Buchner. 2020. ``UltraNest.jl.''
\emph{GitHub Repository}. GitHub.
\url{https://github.com/bat/UltraNest.jl}.

\leavevmode\hypertarget{ref-Skilling2004}{}%
Skilling, J. 2004. ``Nested Sampling.'' Edited by Roland Preuss Rainer
Fischer and Udo von Toussaint. \emph{AIP Conference Proceedings} 735
(1): 395. \url{https://doi.org/10.1063/1.1835238}.

\end{document}